\documentclass[12pt]{article}
\usepackage{amsmath, amssymb, amsthm, float, fullpage, graphicx, multirow,  multicol, parskip,   setspace, subcaption}
\usepackage{tikz}

\usepackage{hyperref}
\usepackage{natbib}

\definecolor{PennRed}{RGB}{152, 30 50}
\definecolor{PennBlue}{RGB}{0, 44, 119}
\definecolor{PennGreen}{RGB}{94, 179,70}
\definecolor{PennViolet}{RGB}{141, 76, 145}
\definecolor{PennSkyBlue}{RGB}{14, 118, 188}
\definecolor{PennOrange}{RGB}{243, 117, 58}
\definecolor{PennBrightRed}{RGB}{223,82, 78}

\hypersetup{pdfborder = {0 0 0.5 [3 3]}, colorlinks = true, linkcolor = PennBrightRed, citecolor = PennSkyBlue}

\title{Protocol for an Observational Study of the Association of High School Football Participation on Health in Late Adulthood}
\author{Timothy G. Gaulton\thanks{Perelman School of Medicine, University of Pennsylvania.}, Sameer K. Deshpande\thanks{Computer Science and Artificial Intelligence Laboratory, Massachusetts Institute of Technology.}, Dylan S. Small\thanks{The Wharton School, University of Pennsylvania.}, Mark D. Neuman\footnotemark[1]}

\begin{document}
\maketitle

\def\C{\mathbb{C}}
\def\R{\mathbb{R}}
\def\Q{\mathbb{Q}}
\def\Z{\mathbb{Z}}
\def\N{\mathbb{N}}

\def\P{\mathbb{P}}
\def\E{\mathbb{E}}

\def\bR{\mathbf{R}}
\def\bZ{\mathbf{Z}}
\def\bX{\mathbf{X}}
\def\br{\mathbf{r}}
\def\bx{\mathbf{x}}

\begin{abstract}
\singlespacing{

\noindent \textbf{Background}: 
American football is the most popular high school sport and is among the leading cause of injury among adolescents. 
While there has been considerable recent attention on the link between football and cognitive decline, there is also evidence of higher than expected rates of pain, obesity, and lower quality of life among former professional players, either as a result of repetitive head injury or through different mechanisms. 
Previously hidden downstream effects of playing football may have far-reaching public health implications for participants in youth and high school football programs.
\vspace{0.1in}
\noindent \textbf{Methods and Analysis}: Our proposed study is a retrospective observational study that compares 1,153 high school males who played varsity football with 2,751 male students who did not. 1,951 of the control subjects did not play any sport and the remaining 800 controls played a non-contact sport. Our primary outcome is self-rated health measured at age 65. To control for potential confounders, we adjust for pre-exposure covariates with matching and model-based covariance adjustment. We will conduct an ordered testing procedure designed to use the full pool of 2,751 controls while also controlling for possible unmeasured differences between students who played sports and those who did not. We will quantitatively assess the sensitivity of the results to potential unmeasured confounding. The study will also assess secondary outcomes of pain, difficulty with activities of daily living, and obesity, as these are both important to individual well-being and have public health relevance.

\vspace{0.1in}
}
\end{abstract}

\newpage
\section{Background and Motivation}
\label{sec:background_motivation}

While the number of students playing high school football in the United Status (US) has decreased over the past decade, it still remains the most popular sport among high school boys with over 1 million participants between 2017 and 2018 \citep{NFSHSA2017}
It is estimated that there are over 440,000 injuries that require medical attention related to participation in high school football in competition or in practice \citep{Comstock2017}
Most injuries are strains or contusions to the knee and ankle \citep{Badgeley2013, TacklingYouthFB2015}
 Injuries to that head such as a concussion represent 9-11\% of all injuries, although these are more frequent in football than other sports \citep{Badgeley2013, TacklingYouthFB2015} and are often not reported \citep{McCrea2004}.
The rate of injury has led to public concerns about the immediate and potential long-term health consequences of playing football, even in adolescence.  
And while there has been an appropriate awareness and research on chronic traumatic encephalopathy and cognitive decline, there may be other more prevalent effects such as pain, obesity, and poor quality of life that result from participation in football.

There is an absence of published data on the long-term health outcomes of playing high school football.  
Studies have either been cross-sectional \citep{Laurson2007} or restricted to outcomes on cognition and behavior \citep{Alosco2017}.
There is far more data on the health consequences of playing professional football and while the sport is highly visible, the prevalence of participation dwarfs in comparison to the number of athletes who participate in high school football.  
It is well known that playing professional football poses significant risk on current and future health \citep{Belson2018, Belson2019}
In part due to the physical nature of the sport, the average body mass index (BMI) of football athletes tends to be higher than age-matched controls \citep{Harp2005}.
This is also true for high school athletes \citep{Laurson2007}. 
The excess weight needed to compete in football has been shown to be associated with an increased risk of chronic pain, cardiometabolic disease, and sleep disordered breathing in later life \citep{Churchill2018}.
Other cardiovascular risk for factors such as hypertension are more prevalent in professional athletes even in those who were still playing and even regardless of their position on the field \citep{Tucker2009}.  

While it is unclear whether the same concerns apply to high school football, the sheer number of participants alone mandates that the long-term health consequences be established.  
The importance of defining health outcomes is further demonstrated by the fact that events in early life, positive or negative, can have a significant effect on health in adulthood. 
Traumatic brain injury prior to the age of 25 is associated with an increased probability of receiving a disability pension and welfare benefits \citep{Sariaslan2016}.
In a study of young adults, the risk of self-reported knee osteoarthritis was 5 times higher in those with a history of knee injury compared to those that did not report a knee injury \citep{Gelber2000}.
The potential dangers of football, particularly as they relate to long-term health, has led to recent rule changes, like moving the kickoff line forward to reduce returns, that have decreased the annual rate of concussions \citep{Wiebe2018}.
However, as we lack a complete understanding of how football affects health, substantial risk may remain and further interventions are likely still needed.  

In this light, we choose to focus on a set of health outcomes that encompassed individual well being and the ability to maintain independence in later life as well as outcomes that are important to public health.  
Our primary outcome will be self-rated health, a simple and universal measure of health.  
We will also assess pain, functional impairment, and obesity.  
These conditions are highly prevalent in adulthood and represent some of the biggest public health issues in the US

In principle, the ``gold standard'' approach to determine the causal relationship between playing high school football and health in later life would randomize high-school students to two groups, one assigned to play football and one assigned to not play football, and then compare their health in adulthood. 
This approach is infeasible and therefore, requires techniques using observational data.  
Unfortunately, treated and control groups can often differ in important characteristics prior to treatment and observed differences in outcomes between two groups may be attributable to differences in baseline characteristics rather than the treatment itself.

We therefore will conduct a matched observational study, in which we divide the football players and controls into smaller subgroups which are relatively homogeneous along a range of baseline covariates. We then compare the outcomes within each matched set, after adjusting for residual imbalances in the distribution of these baseline covariates between the football players and controls.

We will use data from the Wisconsin Longitudinal Study (WLS) \citep{Herd2014}, a a long-term study of a random sample of 10,317 men and women who graduated from Wisconsin high schools in 1957 to determine the association between participation in high-school football and self-rated health, pain, and obesity in late adulthood. 
The WLS is an optimal observational dataset to answer these hypotheses.  
WLS captures early-life exposures that are important predictors of health and well-being in later life such parental socioeconomic status, occupation, and education level, family structure, and race.  
It also captures whether study participants participated in high school football and contains detailed measurements of their health in adulthood.  
WLS provides a robust and longitudinal dataset, overcoming the limitations of prior cross-sectional studies, to compare the health in later adulthood of those who played high school football to those who did not, after carefully controlling for a potential confounders.

\section{Methods}
\label{sec:methods}

\subsection{Study Population}
\label{sec:study_population}

The WLS randomly samples 10,317 high school graduates in 1957 (one-third of all graduates) and has observed them since then. 
For more details about the WLS, please see \citet{Herd2014}.
We restricted out attention to an initial pool of 4,991 men (48.38\%). 
WLS researchers performed systematic reviews of high school yearbooks between January 2005 and February 2011, recording, among other things, sports participation data. 
Table~\ref{tab:sports_participation} reports the number of participants in each sport recorded by the WLS. We excluded the 1,221 men (24.46\%) for whom yearbook information is not available as well as an additional 245 men for whom yearbook information was available but sport participation information was unavailable. 
We further excluded men who participated in non-football collision sports (i.e. hockey and wrestling) since these sports also involve frequent, violent collisions with other players or the ground and our goal is to isolate the effect of playing football specifically. 
We also excluded men who participated in a sport listed as ``Other.''
In total, we excluded 170 men on the basis of their sports participation.
Following these initial exclusions, we are left with an pool of 3,355 subjects eligible for our study. 
Of the eligible subjects, 1,259 (37.53\%) participated in football and 2,096 (62.47\%) did not. 

\begin{table}[H]

\centering
\caption{Number of participants in each sport recorded by the WLS. The column labelled Initial Pool records the number of participations in each sport before any exclusions}
\label{tab:sports_participation}
\begin{tabular}{lrrr}
\hline
Sport & Initial Pool & Eligible Football Players & Eligible Sport Controls \\ \hline
Baseball & 786 & 453 & 273 \\
Basketball & 1046 & 654 & 328 \\
Cross Country & 84 & 13 & 54 \\
Curling & 36 & 10 & 23 \\
Football & 1420 & 1259 & 0 \\
Track & 804 & 468 & 238 \\
Volleyball & 127 & 69 & 45 \\
Wrestling & 130 & 0 & 0 \\
Swimming & 60 & 14 & 38 \\
Hockey & 11 & 0 & 0 \\
Tennis & 73 & 32 & 39 \\
Other & 34 & 0 & 0 \\ \hline
\end{tabular}

\end{table}

A key advantage to using the WLS is the availability of several baseline variables measured when the subjects were still in high school.
Table~\ref{tab:baseline_covariates} lists several of these covariates which may be potential confounders and which we adjusted for in our analysis.
For example, WLS captures family socioeconomic characteristics such as parental income, education and occupation.
Prior research has shown that children from a disadvantaged background are more likely to develop chronic disease and have functional limitations in adulthood compared to children from higher socioeconomic classes \citep{Haas2008}.
Furthermore, negative health events like infections and poor overall health in childhood have been associated with lower self-rated health and increased disability in adulthood \citep{Blackwell2001, Haas2007}.

\begin{table}[H]
\centering
\singlespacing
\small
\caption{Description of potential confounders considered}
\label{tab:baseline_covariates}

\begin{tabular}{lp{14cm}}
\hline
WLS Id & Description  \\ \hline
srbmi & Standardized BMI  \\
ixc01rer & Self-rating of childhood health \\
ixc13rer & Ever missed 1+ month of school due to health  \\
ixc14rer & Ever confined to bed or home for 1+ month due to health \\
ixc15rer & Sports or physical activity ever restricted for 3+ months due to health \\
ixc02rer & Had asthma as a child \\ 
ixc07rer & Had polio as a child  \\
ixc10rer & Had pneumonia as a child \\
ixc12rer & Had infectious mononucleosis as a child \\
ixc03rer & Had frequent ear infections as a child  \\
ixc04rer & Had tonsils and/or adenoids removed as a child  \\
ixc05rer & Had chronic bronchitis as a child  \\
ixc06rer & Had pertussia as a child  \\
ixc08rer & Had diptheria as a child  \\
ixc09rer & Had hepatitis as a child  \\
ixc11rer & Had meningitis as a child  \\
ixa06rec & Any parent or sibling had heart attack before age 55  \\
ixa07rec & Any parent or sibling had heart attack after age 55  \\
ixa04rec & Any parent or sibling had stroke before age 65 \\
ixa02rec & Any parent or sibling had high blood pressure  \\
ixa03rec & Any parent or sibling had high blood cholesterol  \\
ixa08rec & Any parent or sibling had diabetes \\
ixa09rec & Any parent or sibling had Alzheimer's \\
ixa11rec & Any parent or sibling had osteoperosis  \\
ixt06rer & Subject smoked regularly before age 16  \\
ixt02rer & Other household members smoked  \\
ixt03rer & Did mother smoke  \\
ixt04rer & Did father smoke \\ 
ixt05rer & Did any other household members besides parents smoke  \\
bklvpr & Lived with both parents  \\
wrmo57 & Mother had job in 1957  \\
ocsf57 & Occupational prestige of father's job in 1957  \\
bmfaedu & Father's education \\
bmpin1 & Parental income in 1957  \\
hsdm57 & Denominational control of high school  \\
hsrankq & High school rank  \\
zpedyr & Planned years of further education \\
spocasp3 & Occupational prestige of job aspiration \\
zfrplc & Friends planned to attend college  \\ 
tchcntq & Discussed plans with teachers  \\
parcntq & Discussed plans with parents  \\
parencq & Parents' thoughts about attending college  \\ \hline
\end{tabular}

\end{table}

\subsection{Primary and Secondary Outcomes}
\label{sec:outcomes}

Our primary outcome is a subject's self-rated health at age 65.  
In 2004, the WLS conducted phone interviews with enrolled subjects and asked the following question:  ``In general, would you say your health is excellent, very good, good, fair, or poor?''
Self-rated health is a widely used and holistic indicator of health \citep{Idler1997}.
It correlates strongly with mortality even after adjustment for more objective measures of health \citep{Mossey1982}.
In line with prior work, we dichotomized subjects' responses, coding ``excellent'', ``very good'', and ``good'' as 1 and coding ``fair'' and ``poor'' as 0 \citep{Idler1997}.

For secondary outcomes, we selected WLS questions that captured important patient-centered measures of health and quality of life; these included questions on pain, functional impairment, and subject weight.  
These outcomes were also recorded in the 2004 WLS interviews.  
As there is no standard assessment of pain, we used the following question from WLS: ``during the past 4 weeks, how many of your activities were limited by pain or discomfort?''
We coded subject responses as 1 if they responded that some, most, or all of their activities were limited and 0 if they responded that none or a few activities were limited.  For functional impairment, we assessed whether subjects had difficulty completing activities of daily living (ADL). 
Maintaining independence in performing ADLs is a central aspect of functioning and well-being but impairment is common with older age \citep{Hardy2004}.
WLS asked subjects: ``During the past 4 weeks, have you been able to eat, bathe, dress and use the toilet without difficulty?'' coded as 0 for no difficulty and 1 for any difficulty.  
Finally, we determined maximum lifetime weight using the 2004 WLS mail survey in 2004.  
Subjects reported their maximum lifetime weight and the age at which they weighed the most. 
From these responses, we computed the maximum adult BMI for each subject who reported that they reached their maximum weight when they were 18 years old or older.  
We choose to use maximum lifetime weight as an outcome because weight assessed in older age fails to separate subjects who were obese and then lost weight because of illness from subjects whose weight never changed. 
Models that incorporate weight history better predict mortality than models that use a conventional approach of using weight from a single time point \citep{Stokes2016}.  

We also assessed whether a subject indicated having a diagnosis of cancer at the 2004 WLS interview.  
As there is no scientific basis for a connection between football and cancer, this outcome tests the validity of our match and eventual study findings.  
We excluded cancers that have are known to have a strong association with smoking such as lung and bladder cancer as subjects who played football may be less likely to smoke.   
WLS codes the type of cancer using the International Classification of Diseases, Ninth Revision (ICD-9).  
We then created a binary outcome indicating whether or not a subject had cancer.

Table~\ref{tab:missing_outcomes} summarizes the total number of eligible subjects who are missing each outcome considered.
We dropped the 334 football players and 573 controls missing the primary outcome from our analysis.
This represents a 27.1\% reduction in our sample size.

\begin{table}[H]
\centering
\footnotesize
\caption{Number (proportion) of eligible subjects missing each outcome}
\label{tab:missing_outcomes}
\begin{tabular}{lrrrr}
\hline
Outcome & Total & FB & Sport Controls & Non-sport Controls \\ \hline
Self-rated Health & 909 (27.09\%) & 334 (26.53\%) &176 (24.24\%) &399 (29.12\%) \\
Pain Limits Activities & 2487 (74.13\%) & 933 (74.11\%) & 529 (72.87\%) & 1025 (74.82\%) \\
ADL Difficulty & 1044 (31.12\%) & 377 (29.94\%) & 218 (30.03\%) & 449 (32.77\%)\\
Cancer & 1045 (31.15\%) & 378 (30.02\%) & 218 (30.03\%) & 449 (32.77\%) \\
Max Adult BMI&1416 (42.21\%) & 502 (39.87\%) & 306 (42.15\%) & 608 (44.38\%) \\
\end{tabular}

\end{table}

To examine whether playing high school football increased the likelihood of attrition from the WLS or the availability of outcomes at age 65, we will perform an attrition analysis and fit logistic regression models to estimate the availability of each outcome using all of the baseline covariates listed in Table~\ref{tab:baseline_covariates}.

\subsection{Matching methodology}
\label{sec:matching}

Randomizing treatment assignment is well-known to \textit{balance} the distribution of covariates, both measured and unmeasured, between treated and control groups.
In this setting, a direct comparison of each group's average outcome is a rather reasonable causal estimate.
When treatment is not randomly assigned, however, there tends to be substantial covariate imbalance between treated and control groups and such a comparison may be confounded.
Matching is a popular tool to overcome this hurdle by partitioning the study population into several matched sets consisting of one treated subject and one or more control subjects or one control subject and one or more treated subjects. 
These sets are constructed in such a way that the matched sets are aligned along all of the observed covariates. 
In a sense, matching is a way to sets of comparable subjects. 
For a comprehensive review on the role of matching in observational studies, see \citet{Stuart2010}.

In our planned study, we construct matched sets containing one treated subject and a variable number of controls so that the distribution of the covariates listed in Table are as similar as possible within matched sets. 
To do so, we begin by estimating the propensity score, or the conditional probability of being assigned treatment given covariates. 
We also compute a distance matrix containing the rank-based Mahalanobis distance between each pair of treated and control subjects' covariates. We then add a propensity-score caliper to this distance matrix. The combination of a propensity-score caliper and rank-based Mahalanobis distance matrix strives to achieve a compromise between overall covariate balance and closeness of covariates of matched subjects \citep{GuRosenbaum1993}.

We then perform variable ratio matching \citep{Ming2000, Pimentel2015} using the following procedure: given an integer $K \geq 2$ for each integer $k = 1, \ldots, K$ we identify subjects whose propensity scores fall within the interval $\left(1/k+1, 1/k\right]$ (with suitable endpoints adjustment for $k = 1$ and $k = K$) and perform 1:k matching using the \textbf{optmatch} \citep{Hansen2006} package in \texttt{R} \citep{R2018}. 
Within each of these blocks, if there are ever more than $k$ times as many control subjects as treated subjects, some control subjects are dropped from the analysis. Moreover, when there are more treated subjects than control subjects in the selected set, we build pair matches, optimally discarding the extra treated subjects who are most dissimilar from the controls under consideration \citep{Rosenbaum2012}. 
This matching procedure does not generally use the full set of study subjects; instead it attempts to optimally discard subjects for whom there are not sufficiently comparable subjects with the opposite treatment assignment.

Since there are different patterns of outcome missingness, we first stratify subjects into different groups based on which outcomes were available for them. 
We then run the above procedure separately within each stratum.
This way, subjects are only matched to subjects who have the same available outcomes.
To determine $K$, the maximum number of controls that we allow to be matched to a single football player, we run our matching procedure over a range of possible $K$ values (in this case 2 to 8) and select the $K$ which uses the largest number of study subjects and which adequately balances covariates (i.e. absolute standardized differences less than 0.2 SDs) between (i) all matched football players and controls and (ii) all matched football players and controls with each secondary outcome available.

Following \citep{Deshpande2017}, we also considered two subgroups as alternative control groups: those controls who played a sport with low-incidence of head trauma like basketball or tennis (sport controls) and those controls who did not play any sport at all (non-sport controls). 
It may be the case that these two subgroups differ along unmeasured dimensions related to personality or fitness that may affect our study outcomes. 
A convincing study of an effect of playing football specifically would show consistent evidence across comparisons of football players with all controls, sport controls, and non-sport controls. 
Moreover, comparability of the two subgroups of controls would mitigate concern about these unmeasured confounders. 
In all, we perform four comparisons -- football vs all controls, football vs sport-controls, football vs non-sport controls, and sport controls vs non-sport controls. We construct a separate match for each comparison.

\subsection{Ordered Hypothesis Testing}
\label{sec:ordered_testing}

After matching, we fit conditional logistic regression models to estimate the effect of playing football on each binary outcome and report effects as odds ratios (ORs). 
For the continuous secondary outcome, maximal adult BMI, we fit a linear regression model with matched set indicators and estimate the effect of playing football by its estimated regression coefficient.
We assess effect sizes for the continuous outcomes using the popular criteria of \citet{Cohen1988} and \citet{Sawilowsky2009}: between 0.01 and 0.2 SDs for very small effects, between 0.2 and 0.5 SDs for small effects, between 0.5 and 0.8 SDs for medium effects, between 0.8 and 1.2 SDs for large effects, and over 1.2 SDs for very large effects. 
For the binary outcomes, we use the criterion of \citet{Chen2010}; 1.5 for small effect sizes and 5.0 for large effect sizes on the OR scale.

To perform the aforementioned comparisons with different control groups without losing power due to multiple testing, we used the same ordered testing procedure of \citep{Deshpande2017}. 
In that procedure, we test whether the effect of playing football is equal to a specific $\tau_{0}$  using the two alternative control groups only if $\tau_{0}$ falls outside the 95\% confidence interval from the main comparison between football players and all controls. 
For the sake of completeness, we also report results from each comparison even if it is not reached in the ordered testing procedure. 
In such cases, the confidence intervals are left unadjusted for multiple testing and are designated as such.

\section{Results}
\label{sec:results}

Table~\ref{tab:match1} shows a subset of standardized differences from matching football players to all controls.
Prior to matching, compared to all controls, football players had higher standardized BMIs, tended to come from smaller high schools, and were less likely to have smoked regularly before age 16 than all controls.
They were further more likely to rate their overall adolescent health as ``excellent'' and less likely to rate it as merely ``good.''
The imbalance between the football players and all controls along these directions was rather high, with standardized differences exceeding 0.2 SDs in absolute value.
After matching, we find that the standardized differences are all acceptably small.
It is reassuring to see that before and after matching, football players and all controls reported comparable rates of childhood asthma, pneumonia, bronchitis, and family history of high blood pressure, high cholesterol, stroke, heart attacks, and diabetes. 
The other matches were similar (see Tables~\ref{tab:match2} --\ref{tab:match4} in the Appendix).

\begin{table}[H]
\centering
\footnotesize
\caption{Comparison of average baseline variables for football players (FB) and all controls (AC), before and after matching. Standardized differences (Std. Diff.) less than 0.2 in magnitude indicate adequate balance. Before matching, control values are unweighted. After matching, control values are weighted according to the composition of the matched sets (Table~\ref{tab:matched_set_composition} in the Appendix).}
\label{tab:match1}

\begin{tabular}{lrrrrrr} 
\hline
~ & \multicolumn{2}{c}{Before Matching} & \multicolumn{2}{c}{After Matching} & \multicolumn{2}{c}{Std. Diff.} \\
Variable & FB & AC & FB & AC & Before & After \\ \hline
Standardized BMI & 0.199 & -0.134 & 0.123 & 0.057 & \textbf{0.416} & 0.083 \\
Smoking exposure & ~ & ~ & ~ & ~ & ~ & ~ \\
\hspace{1em} Regularly smoked before age 16 (\%) &  30.9 & 41.15 & 32.51 & 34.05  & \textbf{-0.215} & -0.032 \\
\hspace{1em} Mother smoked (\%) & 37.5  & 30.86 & 35.61 & 32.76 & 0.140 & 0.06 \\
\hspace{1em} Father smoked & 90.1 & 90.6 & 90.14 & 89.94 & -0.017 & 0.007 \\
Childhood health & ~ & ~ & ~ & ~ & ~ & ~ \\
\hspace{1em} Self-rating& ~ & ~ & ~ & ~ & ~ & ~ \\
\hspace{2em} Poor (\%) & 0.26 & 0.57 & 0.29 & 0.29 & -0.049 & 0 \\
\hspace{2em} Good (\%) & 7.59 & 14.4& 8.11 & 9.82 & \textbf{-0.219} & -0.055 \\
\hspace{2em} Excellent (\%) & 57.14& 44.11 & 55.6 & 52.05 & \textbf{0.263} & 0.072 \\
\hspace{1em} Had asthma (\%) & 3.54 & 5.27  & 3.55 & 4.12 & -0.085 & -0.028 \\
\hspace{1em} Had Pneumonia (\%) & 14.75 & 13.78 & 13.89 & 13.26 & 0.028 & 0.018 \\
\hspace{1em} Had chronic bronchitis & 3.53 & 4.15 & 3.41 & 3.35 & -0.032 & 0.003 \\
Family history of  & ~ & ~ & ~ & ~ & ~ & ~ \\
\hspace{1em} Heart attack before age 55 (\%) & 13.26 & 14.26 & 13.54 &13.97 & -0.029 & -0.012 \\
\hspace{1em} Stroke before age 55 (\%) & 10.92  & 12.93 & 11.9 & 12.56 & -0.062 & -0.02 \\
\hspace{1em} High blood pressure (\%) & 51.89 & 51.38 & 53.27 & 53.52 & 0.01 & -0.005 \\
\hspace{1em} High blood cholesterol (\%) & 30.82 & 30.11 & 31.1 & 31.22 & 0.015 & -0.002 \\
\hspace{1em} Diabetes (\%) & 32.77 & 30.19 & 32.14 & 31.85 & 0.055 & 0.006 \\
Plan to join military (\%) & 26.53  & 24.83  & 25.36 & 26.23 & 0.039 & -0.02 \\
High school size & 134.016 & 183.381 & 138.538 & 140.124 & \textbf{-0.405} & -0.013 \\
Adolescent IQ & 102.297 & 102.915 & 102.304 & 102.084 & -0.041 & 0.015 \\ \hline
\end{tabular}

\end{table}

\section{Sensitivity Analysis}
\label{sec:sensitivity_analysis}

Controlling for observed baseline covariates through matching is designed to eliminate bias in treatment assignment by balancing the distribution of observable potential confounders between the treated and control groups. 
However, it cannot ensure balance of unobserved confounders unless they are highly correlated with observed confounders. 
Conditional on the full matching, let $\Gamma$ bound the odds ratio of treatment for any pair in a matched set. 
For each marginally significant outcome in both the primary and secondary analysis we will report the $\Gamma$ at which the result is sensitive (i.e. the $\Gamma$ at which the result becomes insignificant). 
This is known as a sensitivity analysis \citep{Rosenbaum2002} and allows us to quantitatively assess how sensitive the results are to bias in the treatment assignment due to the observational nature of the study.  
Larger values of $\Gamma$ provide greater evidence for the study's causal conclusions. 
Sensitivity analyses for the continuous primary and secondary outcomes will be performed with the \texttt{sensitivitymv} package in \texttt{R} where we will use the Huber-Maritz M-test with no trimming at level $\alpha =.05$  
For the binary outcomes, we will use a sensitivity analysis for testing the null hypothesis of no treatment effect using the Mantel-Haenszel test \citep[Section 4.2]{Rosenbaum2002}. 
For the continuous secondary outcome (maximum adult BMI), the sensitivity analysis will be performed on the residuals after regressing the outcomes on the covariates \citep[see, e.g.,][]{Rosenbaum2002a}. 
For the binary secondary outcomes, we will use a sensitivity analysis for testing the null hypothesis of no treatment effect using the Mantel-Haenszel test \citep[Section 4.2]{Rosenbaum2002}.

\newpage
\appendix
\section{Additional Tables}
\label{app:additional_tables}
\begin{table}[H]
\centering
\caption{Composition of the matched sets constructed for each comparison. The first number in the Composition column is the number of exposed subjects (i.e. football players (FB) in the first three columns and sport controls (SC) in the last column) and the second number is the number of controls.}
\label{tab:matched_set_composition}
\begin{tabular}{lrrrr}
\hline
Composition & FB vs AC & FB vs SC & FB vs NSC & SC vs NSC \\  \hline
1:1 & 580 & 482 & 494 & 406 \\
1:2 & 152 & 12 & 67 & 64 \\
1:3 & 49 & 0 & 26 & 8 \\
1:4 & 4 & 0 & 5 & 11 \\
1:5 & 13 & 0 & 2 & 4 \\
1:6 & 1 & 0 & 3 & 1 \\
1:7 & 14 & 0 & 8 & 11 \\ \hline
\end{tabular}
\end{table}

\begin{table}[H]
\centering
\footnotesize
\caption{Comparison of average baseline variables for football players (FB) and sport controls (SC), before and after matching. Standardized differences (Std. Diff.) less than 0.2 in magnitude indicate adequate balance. Before matching, control values are unweighted. After matching, control values are weighted according to the composition of the matched sets (Table~\ref{tab:matched_set_composition}).}
\label{tab:match2}

\begin{tabular}{lrrrrrr} 
\hline
~ & \multicolumn{2}{c}{Before Matching} & \multicolumn{2}{c}{After Matching} & \multicolumn{2}{c}{Std. Diff.} \\
Variable & FB & SC & FB & SC & Before & After \\ \hline
Standardized BMI &0.199 & -0.194 & -0.091 & -0.13 & \textbf{0.514} & 0.052 \\
Smoking exposure & ~ & ~ & ~ & ~ & ~ & ~ \\
\hspace{1em} Regularly smoked before age 16 (\%) & 30.9 & 38.49 & 33.76 & 36.25 & -0.16 & -0.052 \\
\hspace{1em} Mother smoked (\%) & 37.5 & 33.76 & 29.72 & 33.68  & 0.078 & -0.083 \\
\hspace{1em} Father smoked &  90.1 & 88.54 & 90.21 & 88.83 & 0.051 & 0.045\\
Childhood health & ~ & ~ & ~ & ~ & ~ & ~ \\
\hspace{1em} Self-rating& ~ & ~ & ~ & ~ & ~ & ~ \\
\hspace{2em} Poor (\%) & 0.26 & 0  & 0.51 & 0 & 0.072 & 0.142\\
\hspace{2em} Good (\%) & 7.59 & 11.55 & 8.42 & 10.87 & -0.135 & -0.084\\
\hspace{2em} Excellent (\%) & 57.14  & 48.27 & 54.34 & 49.94 & 0.178 & 0.088\\
\hspace{1em} Had asthma (\%) & 3.54 & 5.81 & 2.36 & 5.03 & -0.108 & -0.126\\
\hspace{1em} Had Pneumonia (\%) & 14.75  & 13.91 & 13.61 & 13.8 & 0.024 & -0.005 \\
\hspace{1em} Had chronic bronchitis &  3.53 & 4.85  & 2.9 & 3.7 & -0.066 & -0.04\\
Family history of  & ~ & ~ & ~ & ~ & ~ & ~ \\
\hspace{1em} Heart attack before age 55 (\%) & 13.26  & 15.12 & 12.76 & 14.9 & -0.053 & -0.062 \\
\hspace{1em} Stroke before age 55 (\%) &  10.92 & 14.88& 14.03 & 14.9 & -0.118 & -0.026\\
\hspace{1em} High blood pressure (\%) &  51.89  & 52.09 & 52.3 & 51.59 & -0.004 & 0.014\\
\hspace{1em} High blood cholesterol (\%) &30.82 & 28.14 & 26.53 & 28.15 & 0.059 & -0.036 \\
\hspace{1em} Diabetes (\%) &  32.77  & 31.16  & 29.34 & 31.97 & 0.034 & -0.057\\
Plan to join military (\%) &  26.53 & 28.35 & 26.35 & 29.48	 & -0.041 & -0.07\\
High school size & 134.016 & 160.722 & 140.983 & 151.033 & \textbf{-0.219} & -0.082\\
Adolescent IQ &  102.297 & 102.148 & 102.622 & 101.801 & 0.01 & 0.054\\ \hline
\end{tabular}

\end{table}

\begin{table}[H]
\centering
\footnotesize
\caption{Comparison of average baseline variables for football players (FB) and non-sport controls (NSC), before and after matching. Standardized differences (Std. Diff.) less than 0.2 in magnitude indicate adequate balance. Before matching, control values are unweighted. After matching, control values are weighted according to the composition of the matched sets (Table~\ref{tab:matched_set_composition} in the Appendix).}
\label{tab:match3}

\begin{tabular}{lrrrrrr} 
\hline
~ & \multicolumn{2}{c}{Before Matching} & \multicolumn{2}{c}{After Matching} & \multicolumn{2}{c}{Std. Diff.} \\
Variable & FB & NSC & FB & NSC & Before & After \\ \hline
Standardized BMI & 0.199 & -0.101 & 0.096 & 0.049 & \textbf{0.368} & 0.058\\
Smoking exposure & ~ & ~ & ~ & ~ & ~ & ~ \\
\hspace{1em} Regularly smoked before age 16 (\%) & 30.9 & 42.62 & 37.68 & 35.51 & \textbf{-0.245} & 0.045 \\
\hspace{1em} Mother smoked (\%) & 37.5 & 29.32 & 32.12 & 31.03 & 0.174 & 0.023 \\
\hspace{1em} Father smoked & 90.1 & 91.69 & 90.5 & 91.84 & -0.055 & -0.046\\
Childhood health & ~ & ~ & ~ & ~ & ~ & ~ \\
\hspace{1em} Self-rating& ~ & ~ & ~ & ~ & ~ & ~ \\
\hspace{2em} Poor (\%) & 0.26 & 0.89 & 0.41 & 0.43 & -0.084 & -0.003 \\
\hspace{2em} Good (\%) & 7.59 & 15.97 & 9.57 & 12.24 & \textbf{-0.262} & -0.083\\
\hspace{2em} Excellent (\%) & 57.14 & 41.83 & 52.75 & 48.07 & \textbf{0.31} & 0.095 \\
\hspace{1em} Had asthma (\%) & 3.54 & 4.97 & 4.04 & 5.08 & -0.071 & -0.052\\
\hspace{1em} Had Pneumonia (\%) & 14.75 & 13.72 & 13.14 & 13.69 & 0.03 & -0.016\\
\hspace{1em} Had chronic bronchitis & 3.53 & 3.76 & 3.4 & 3.92 & -0.012 & -0.027\\
Family history of  & ~ & ~ & ~ & ~ & ~ & ~ \\
\hspace{1em} Heart attack before age 55 (\%) & 13.26 & 13.78 & 12.58 & 13.38 & -0.015 & -0.023  \\
\hspace{1em} Stroke before age 55 (\%) & 10.92 & 11.83 & 12.16 & 11.14 & -0.029 & 0.032\\
\hspace{1em} High blood pressure (\%) &51.89 & 50.98 & 52.99 & 51.43 & 0.018 & 0.031 \\
\hspace{1em} High blood cholesterol (\%) & 30.82 & 31.21 & 31.13 & 30.54 & -0.008	 & 0.013\\
\hspace{1em} Diabetes (\%) & 32.77 & 29.65 & 31.55 & 30.44 & 0.067 & 0.024 \\
Plan to join military (\%) &  26.53 & 22.89 & 21.91 & 23.59 & 0.085 & -0.039\\
High school size & 134.016 & 195.866 & 150.304 & 155.284 & \textbf{-0.512} & -0.041\\
Adolescent IQ & 102.297 &103.338 & 102.621 & 102.131 & -0.069 & 0.032\\ \hline
\end{tabular}

\end{table}

\begin{table}[H]
\centering
\footnotesize
\caption{Comparison of average baseline variables for sport controls (SC) and non-sport controls (NSC), before and after matching. Standardized differences (Std. Diff.) less than 0.2 in magnitude indicate adequate balance. Before matching, NSC values are unweighted. After matching, NSC values are weighted according to the composition of the matched sets (Table~\ref{tab:matched_set_composition} in the Appendix).}
\label{tab:match4}

\begin{tabular}{lrrrrrr} 
\hline
~ & \multicolumn{2}{c}{Before Matching} & \multicolumn{2}{c}{After Matching} & \multicolumn{2}{c}{Std. Diff.} \\
Variable & SC & NSC & SC & NSC & Before & After \\ \hline
Standardized BMI & -0.194 & -0.101 & -0.192 & -0.179 & -0.118 & -0.016\\
Smoking exposure & ~ & ~ & ~ & ~ & ~ & ~ \\
\hspace{1em} Regularly smoked before age 16 (\%) & 38.49 & 42.62 & 37.85 & 37.38 & -0.084 & 0.01 \\
\hspace{1em} Mother smoked (\%) & 33.76 & 29.32 & 32.78 & 31.41 & 0.095 & 0.029 \\
\hspace{1em} Father smoked & 88.54 & 91.69 & 88.96 & 90.54 & -0.106 & -0.053\\
Childhood health & ~ & ~ & ~ & ~ & ~ & ~ \\
\hspace{1em} Self-rating& ~ & ~ & ~ & ~ & ~ & ~ \\
\hspace{2em} Poor (\%) & 0 & 0.89 & 0 & 0.49 & -0.134 & -0.075 \\
\hspace{2em} Good (\%) & 11.55 & 15.97 & 11.98 & 13.52 & -0.129 & -0.045\\
\hspace{2em} Excellent (\%) & 48.27 & 41.83 & 47.68 & 45.64 & 0.13 & 0.041 \\
\hspace{1em} Had asthma (\%) & 5.81 & 4.97 & 5.61 & 4.7 & 0.037 & 0.04\\
\hspace{1em} Had Pneumonia (\%) & 13.91 & 13.72 & 13.71 & 13.75 & 0.006 & -0.001\\
\hspace{1em} Had chronic bronchitis & 4.85 & 3.76 & 4.86 & 4.12 & 0.054	 & 0.037\\
Family history of  & ~ & ~ & ~ & ~ & ~ & ~ \\
\hspace{1em} Heart attack before age 55 (\%) &  15.12 & 13.78& 14.53 & 13.95 & 0.038 & 0.017 \\
\hspace{1em} Stroke before age 55 (\%) & 14.88 & 11.83 & 14.29 & 11.72 & 0.09 & 0.075\\
\hspace{1em} High blood pressure (\%) & 52.09 & 50.98 & 51.48 & 49.86 & 0.022 & 0.032\\
\hspace{1em} High blood cholesterol (\%) & 28.14 & 31.21 & 28.33 & 28.19 & -0.067	 & 0.003 \\
\hspace{1em} Diabetes (\%) & 31.16 & 29.65 & 30.54 & 28.97 & 0.033 & 0.034\\
Plan to join military (\%) & 28.35 & 22.89 & 27.46 & 26.66 & 0.125 & 0.018 \\
High school size & 160.722 & 195.866 & 163.449 & 172.503 & -0.263 & -0.068 \\
Adolescent IQ &102.148 & 103.338 & 102.317 & 103.076 & -0.076 & -0.049 \\ \hline
\end{tabular}

\end{table}

\newpage
\bibliography{wls_refs}

\begin{thebibliography}{}

\bibitem[Alosco et~al., 2017]{Alosco2017}
Alosco, M.~L., Kasimis, A.~B., Stamm, J.~M., Chua, A.~S., Baugh, C.~M.,
  Daneshvar, D.~H., Robbins, C.~A., Mariani, M., Hayden, J., Conneely, S., Au,
  R., Torres, A., McClean, M.~D., McKee, A.~C., Cantu, R.~C., Mez, J.,
  Nowinski, C.~J., Martin, B.~M., Chaisson, C.~E., Tripodis, Y., and Stern,
  R.~A. (2017).
\newblock {Age of first exposure to American football and long-term
  neuropsychiatric and cognitive outcomes}.
\newblock {\em Translational Psychiatry}, 7:e1236.

\bibitem[Badgeley et~al., 2013]{Badgeley2013}
Badgeley, M.~A., Mcilvain, N.~M., Yard, E.~E., Fields, S.~K., and Comstock,
  R.~D. (2013).
\newblock {Epidemiology of 10,000 high school football injuries: patterns of
  injury by position played}.
\newblock {\em J Phys Act Health}, 10(2):160--169.

\bibitem[Belson, 2018]{Belson2018}
Belson, K. (2018).
\newblock {A Football Player's Descent Into Pain and Paranoia}.

\bibitem[Belson, 2019]{Belson2019}
Belson, K. (2019).
\newblock {The N.F.L's Obesity Scourge}.

\bibitem[Blackwell et~al., 2001]{Blackwell2001}
Blackwell, D.~L., Hayward, M.~D., and Crimmins, E.~M. (2001).
\newblock Does childhood health affect chronic morbidity in later life?
\newblock {\em Social Science and Medicine}, 52(8):1269--1284.

\bibitem[Chen et~al., 2010]{Chen2010}
Chen, H., Cohen, P., and Chen, S. (2010).
\newblock {How big is a big odds ratio? Interpreting the magnitudes of odds
  ratios in epidemiological studies}.
\newblock {\em Communications in Statistics: Simulation and Computation},
  39(4):860--864.

\bibitem[Churchill et~al., 2018]{Churchill2018}
Churchill, T.~W., Krishnana, S., Weisskopf, M., and et~al. (2018).
\newblock Weight gain and health afflication among former national footbal
  league players.
\newblock {\em American Journal of Medicine}, 131(12):1491--1498.

\bibitem[Cohen, 1988]{Cohen1988}
Cohen, J. (1988).
\newblock {\em {Statistical power analysis for the behavioral sciences}}.
\newblock Erlbaum, Hillsdale, NJ, 2 edition.

\bibitem[Comstock et~al., 2017]{Comstock2017}
Comstock, R., Pierpont, L.~A., Erkenbeck, A.~N., and Bihl, J. (2017).
\newblock {Summary Report: National High School Sports-Related Injury
  Surveillance Study. 2015 - 2016 School Year}.
\newblock Technical report, Colorado School of Public Health.

\bibitem[{Council on Sports Medicine and Fitness}, 2015]{TacklingYouthFB2015}
{Council on Sports Medicine and Fitness} (2015).
\newblock {Tackling in Youth Football}.
\newblock {\em Pediatrics}, 136(5):e1419--e1430.

\bibitem[Deshpande et~al., 2017]{Deshpande2017}
Deshpande, S.~K., Hasegawa, R.~B., Rabinowitz, A.~R., Whyte, J., Roan, C.~L.,
  Tabatabaei, A., Baiocchi, M., Karlawish, J.~H., Master, C.~L., and Small,
  D.~S. (2017).
\newblock {Association of playing high school football with cognition and
  mental health later in life}.
\newblock {\em JAMA Neurology}, 74(8):909--918.

\bibitem[Gelber et~al., 2000]{Gelber2000}
Gelber, A.~C., Hochberg, M.~C., Mead, L.~A., Wang, N.-y., Wigley, F.~M., and
  Klag, M.~J. (2000).
\newblock {Joint Injury in Young Adults and Risk for Subsequent Knee and Hip
  Osteoarthritis}.
\newblock {\em Annals of Internal Medicine}, 133(5):321--328.

\bibitem[Haas, 2007]{Haas2007}
Haas, S. (2007).
\newblock The long-term effects of poor childhood health: an assessment and
  application of retrospective reports.
\newblock {\em Demography}, 44(1):113--135.

\bibitem[Haas, 2008]{Haas2008}
Haas, S. (2008).
\newblock Trajectories of functional health: the `long arm' of childhood health
  and socioeconomic factors.
\newblock {\em Social Science and Medicine}, 66(4):849--861.

\bibitem[Hansen and Klopfer, 2006]{Hansen2006}
Hansen, B.~B. and Klopfer, S.~O. (2006).
\newblock Optimal full matching and related designs via network flows.
\newblock {\em Journal of Computational and Graphical Statistics},
  15(3):609--627.

\bibitem[Hardy and Gill, 2004]{Hardy2004}
Hardy, S.~E. and Gill, T.~M. (2004).
\newblock Recovery from disability among community-dwelling older persons.
\newblock {\em JAMA}, 291(13):1596--1602.

\bibitem[Harp and Hecht, 2005]{Harp2005}
Harp, J.~B. and Hecht, L. (2005).
\newblock Obesity in the {National Football Leauge}.
\newblock {\em JAMA}, 293(9):1061--1062.

\bibitem[Herd et~al., 2014]{Herd2014}
Herd, P., Carr, D., and Roan, C. (2014).
\newblock {Cohort profile: Wisconsin longitudinal study (WLS)}.
\newblock {\em International Journal of Epidemiology}, 43(1):34--41.

\bibitem[Idler and Benyamini, 1997]{Idler1997}
Idler, E.~L. and Benyamini, Y. (1997).
\newblock Self-rated health and mortality: a review of twenty-sever community
  studies.
\newblock {\em Journal of Health and Social Behavior}, 38(1):21 -- 37.

\bibitem[Laurson and Eisenmann, 2007]{Laurson2007}
Laurson, K.~R. and Eisenmann, J.~C. (2007).
\newblock Prevalence of overweight among high school football lineman.
\newblock {\em JAMA}, 297(4):363--364.

\bibitem[McCrea et~al., 2004]{McCrea2004}
McCrea, M., Hammeke, T., Olsen, G., Leo, P., and Guskiewicz, K. (2004).
\newblock {Unreported Concussion in High School Football Players: Implications
  for Prevention}.
\newblock {\em Clinical Journal of Sport Medicine}, 14(1):13--17.

\bibitem[Ming and Rosenbaum, 2000]{Ming2000}
Ming, K. and Rosenbaum, P.~R. (2000).
\newblock {Substantial Gains in Bias Reduction from Matching with a Variable
  Number of Controls}.
\newblock {\em Biometrics}, 56(1):118--124.

\bibitem[Mossey and Shapiro, 1982]{Mossey1982}
Mossey, J.~M. and Shapiro, E. (1982).
\newblock Self-rated health: a predictor of mortality among the elderly.
\newblock {\em American Journal of Public Health}, 72(8):800--808.

\bibitem[Pimentel et~al., 2015]{Pimentel2015}
Pimentel, S.~D., Yoon, F., and Keele, L. (2015).
\newblock {Variable-ratio matching with fine balance in a study of the Peer
  Health Exchange}.
\newblock {\em Statistics in Medicine}, 34(30):4070--4082.

\bibitem[{R Core Team}, 2018]{R2018}
{R Core Team} (2018).
\newblock {\em R: A Language and Environment for Statistical Computing}.
\newblock {R Foundation for Statistical Computing}, {Vienna, Austria}.

\bibitem[Rosenbaum, 2002a]{Rosenbaum2002a}
Rosenbaum, P.~R. (2002a).
\newblock {Covariance Adjustment in Randomized Experiments and Observational
  Studies}.
\newblock {\em Statistical Science}, 17(3):286--327.

\bibitem[Rosenbaum, 2002b]{Rosenbaum2002}
Rosenbaum, P.~R. (2002b).
\newblock {\em {Observational Studies}}.
\newblock Spring, New York.

\bibitem[Rosenbaum, 2012]{Rosenbaum2012}
Rosenbaum, P.~R. (2012).
\newblock {Optimal Matching of an Optimally Chosen Subset in Observational
  Studies}.
\newblock {\em Journal of Computational and Graphical Statistics}, 21(1):57 --
  71.

\bibitem[{Sam Gu} and Rosenbaum, 1993]{GuRosenbaum1993}
{Sam Gu}, X. and Rosenbaum, P.~R. (1993).
\newblock {Comparison of Multivariate Matching Methods: Structures, Distances,
  and Algorithms}.
\newblock {\em Journal of Computational and Graphical Statistics},
  2(4):405--420.

\bibitem[Sariaslan et~al., 2016]{Sariaslan2016}
Sariaslan, A., Sharp, D.~J., D'Onofrio, B.~M., Larsson, H., and Fazel, S.
  (2016).
\newblock {Long-Term Outcomes Associated with Traumatic Brain Injury in
  Childhood and Adolescence: A Nationwide Swedish Cohort Study of a Wide Range
  of Medical and Social Outcomes}.
\newblock {\em PLoS Medicine}, 13(8):15--19.

\bibitem[Sawilowsky, 2009]{Sawilowsky2009}
Sawilowsky, S.~S. (2009).
\newblock {New Effect Size Rules of Thumb}.
\newblock {\em Journal of Modern Applied Statistical Methods}, 8(2):597--599.

\bibitem[Stokes and Preston, 2016]{Stokes2016}
Stokes, A. and Preston, S.~H. (2016).
\newblock Revealing the burden of obesity using weight histories.
\newblock {\em Proceedings of the National Academy of Science},
  113(3):572--577.

\bibitem[Stuart, 2010]{Stuart2010}
Stuart, E.~A. (2010).
\newblock {Matching Methods for Causal Inference: A Review and a Look Forward}.
\newblock {\em Statistical Science}.

\bibitem[{The National Federation of State High School Associations},
  2017]{NFSHSA2017}
{The National Federation of State High School Associations} (2017).
\newblock {\em {NFHS Handbook 2017-18}}.

\bibitem[Tucker et~al., 2009]{Tucker2009}
Tucker, A.~M., Vogel, R.~A., Lincoln, A.~E., and et~al. (2009).
\newblock Prevalence of cardiovascular disease risk factors among {National
  Football League} players.
\newblock {\em JAMA}, 301(20):2111--2119.

\bibitem[Wiebe et~al., 2018]{Wiebe2018}
Wiebe, D.~J., D'Alonzo, B.~A., Harris, R., Putukian, M., and Campbell-McGovern,
  C. (2018).
\newblock {Association Between the Experimental Kickoff Rule and Concussion
  Rates in Ivy League Football}.
\newblock {\em JAMA}, pages 2018--2019.

\end{thebibliography}

\end{document}